\documentclass[12pt]{JHEP3}

\usepackage{amsmath,amssymb,graphicx,epsf}

\def\d#1{\partial_{#1}}

\def\x{\times}
\newcommand{\tr}{\hbox{tr}}
\def\BOX{\mathord{\vbox{\hrule\hbox{\vrule\hskip 3pt\vbox{\vskip
3pt\vskip 3pt}\hskip 3pt\vrule}\hrule}\hskip 1pt}}

\newcommand{\RR}{\mathbb{R}}

\title{A World-Volume Perspective on the Recombination of Intersecting
Branes}

\author{J. Erdmenger$^*$, Z. Guralnik$^*$, R. Helling$^\dagger$ and
I. Kirsch$^*$\\

$^*$ Institut f\"ur Physik, Humboldt-Universit\"at zu Berlin, Newtonstra\ss
e
15,\\ D-12489 Berlin, Germany\\

$^\dagger$ Department for Applied Mathematics and
Theoretical Physics, Cambridge University, Wilberforce Road, Cambridge
CW3 0WA, United Kingdom\\

{\tt E-mail: jke@physik.hu-berlin.de, zack@physik.hu-berlin.de,
helling@atdotde.de, ik@physik.hu-berlin.de}}

\abstract{We study brane recombination for supersymmetric configurations
of intersecting branes in terms of the world-volume field theory.
This field theory contains an impurity,  corresponding to the degrees of
freedom localized at the intersection. The Higgs branch, on which the
impurity
fields condense, consists of vacua for which the intersection is
deformed into a smooth calibrated manifold.  We show this explicitly using a
superspace formalism for which the calibration equations arise naturally
from F- and D-flatness.}
\preprint{HU-EP-03/44\\ DAMTP-2004-84\\ hep-th/0309043}

\begin{document}


\section{Introduction}
\setcounter{equation}{0}

It is well known that a supersymmetry preserving set of
intersecting branes may merge to form a single brane on a smooth
calibrated surface (see \cite{Smith} for a review). This is
expected to occur via the condensation of world-volume degrees of
freedom localized at the intersection.  However, the manner in
which the Higgs branch of the world-volume gauge theory describes
smooth calibrated surfaces has not been studied explicity in much
detail.  Filling this gap is the aim of this note. We shall do so
using a superspace for which calibrated geometries on the Higgs
branch arise naturally as solutions of F- and D-flatness
conditions. The gauge theories which we write down are suitable
for describing both the Coulomb branch on which the branes are
separated and the Higgs branch on which the branes have merged on
a calibrated surface.

These gauge theories contain impurities, given by degrees of
freedom constrained to the lower-dimensional subspace where the
branes intersect. It is convenient to describe such a field theory
by a superspace which spans the space-time directions of the
impurity. This superspace may also be used to described the fields
in the ambient space which are not localized at the impurity.  The
formalism in which degrees of freedom in higher dimensions are
described in a lower-dimensional superspace was developed
originally in \cite{MSS,AHGW} and has been applied to study a
number of theories, including the defect conformal theories which
arise for systems of interesecting branes
\cite{EGK,CEGK1,CEGK2,CEGK3}.

In the context of brane intersections preserving eight
supercharges, the holomorphic  curves which arise on the Higgs
branch are solutions of the F- and D-flatness conditions for a
superspace which spans the mutual coordinates of the intersection.
For triple (and quadruple) intersections preserving four
supercharges,  we will find F- and D-flatness corresponding to
special Lagrangian conditions. These conditions will take the form
of the equations of motion of an abelian Chern-Simons theory and a
gauge fixing respectively.

For a particular triple intersection of D6-branes preserving four
supercharges, the Higgs branch solutions of the F- and D-flatness
conditions correspond to the union of a special Lagrangian plane
and a holomorphic curve times a line. This configuration lifts to
a $G_2$ manifold in M-theory which shares many features with
G2-manifolds discussed in \cite{AW,GT}.

There are a number of reasons to be interested in the process of
brane recombination. In the context of intersecting brane-worlds
\cite{BGKL1,BGKL2,AFIRU,IMR}, see also \cite{K1,K2,K3},
the Higgs mechanism is believed to be realized by
brane-recombination \cite{Ibanez}. In this setting intersecting
branes merge via open string tachyon condensation. The
supersymmetric triple D6-brane intersection which we will study
provides a controlled toy model with which to explicitly
demonstrate the brane-world Higgs mechanism.  A precise treatment
of recombination by tachyon condensation has not been given and
would seem to require string field theory, except in certain small
angle approximation \cite{Hashimoto}. Approximate treatments have
been given using effective tachyon field theories
\cite{Hashimoto,Huang,HT,JT}.  For the supersymmetric brane
recombination which we will describe, the field theory treatment
suffices.  Recombination in a supersymmetric case was also
considered in \cite{HT}.  Another reason to be interested in brane
recombination is  that it is closely related (via U-dualities) to
processes such as the brane collision in ekpyrotic
\cite{Khoury:2001wf} or cyclic universes \cite{ST}. In this
context it is useful to have a description which allows to study
the dynamics of transitions between the Coulomb branch, with
separated branes, and the Higgs branch on which the branes have
recombined.

\EPSFIGURE{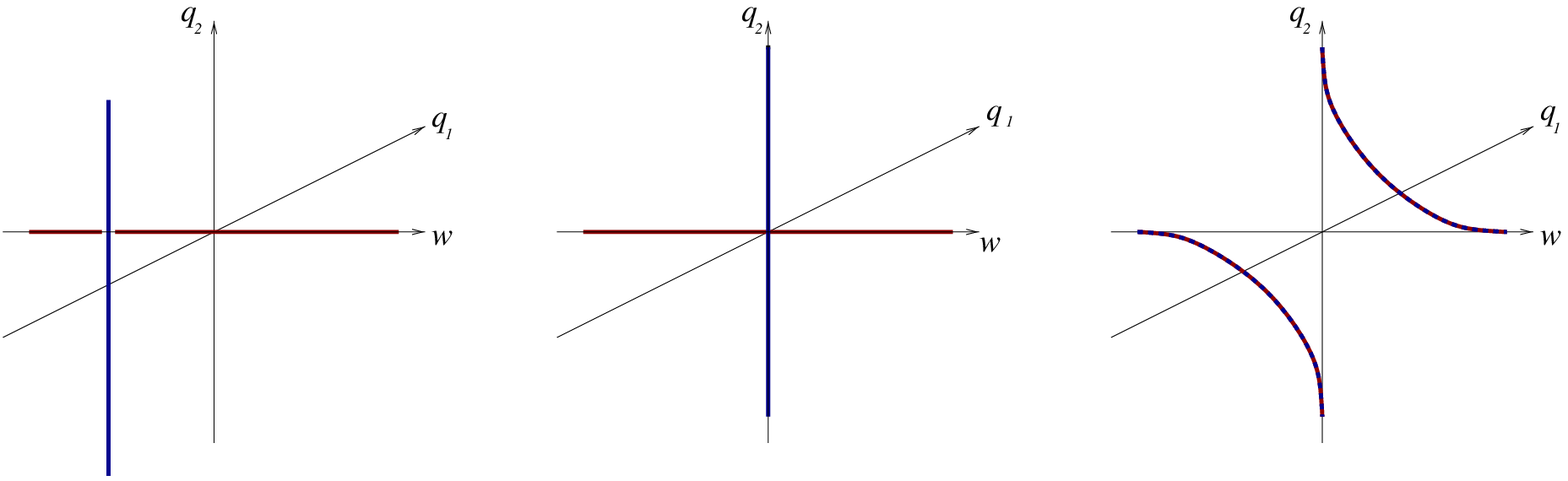,width=\hsize}{Branes touching and the
  intersection being resolved}
\label{resolved}

The organization of this paper is as follows. In section 2, we
discuss intersecting pairs of branes which preserve eight
supercharges and wrap holomorphic curves on the Higgs branch. This
section reviews and expands upon results in \cite{CEGK1}. In
section 3, we consider intersections preserving four supercharges,
for which the F and D-flatness conditions are the the special
Lagrangian conditions. We study the particular example of three
(and four) intersecting D6-branes. Finally, in section $4$, we
numerically compute correlations of fields on the different
asymptotic regions of branes which have recombined into a
holomorphic curve $xy=c$. We find that the correlation vanishes in
the singular limit $c\rightarrow 0$ in which the branes can
separate.

\section{Holomorphic curves from intersecting branes preserving
eight supercharges}

The low energy dynamics of intersecting branes is described by a
field theory with impurities. Some of the earliest studies of such
impurity field theories may be found in \cite{S,GS,KS}.  In many
instances, this theory is a non-trivial defect conformal field
theory \cite{KR,DFO,AdWFK,EGK,CEGK1,CEGK2,CEGK3}), or dCFT. In
this section we review the superspace description of defect field
theories and expand upon some results \cite{CEGK1} concerning the
Higgs branch of the dCFT describing intersecting D3-branes.
Although the world-volume of this dCFT is the singular space $xy =
0$, we will explicitly see that the classical Higgs branch has a
geometric interpretation as the smooth resolution $xy =c$.

Consider two stacks of D3 branes,  one of which spans the
directions $x^{0,1,2,3}$, while the other spans $x^{0,1,4,5}$. The
two stacks are at the origin in the transverse $x^{6,7,8,9}$
directions.
In addition to the ${\cal N}=4$ gauge theory that lives on each
stack of parallel D3-branes, there are additional massless fields
that arise from open strings stretching between the orthogonal
branes. These fields are localized at the 1+1 dimensional
intersection. There is an unbroken $(4,4)$ supersymmetry which
includes translations in the $0,1$ directions.  The $(4,4)$
algebra is a common subalgebra of the two four-dimensional
${\mathcal N} = 4$ algebras associated with each parallel stack of
D3-branes.

Although the theory contains both two- and four-dimensional
degrees of freedom, one can write the action using a
two-dimensional superspace.   The degrees of freedom which
propagate in four dimensions are described by superfields with
continuous indices which parameterize the world-volume directions
transverse to the intersection.  The directions parallel to the
intersection are included in the superspace.   In \cite{CEGK1} the
action for this theory was constructed using two-dimensional
$(2,2)$ superspace.
The formalism which we will use below is trivially generalized to
intersecting D-branes of other dimensions, such as D5-branes
intersecting over four dimensional ${\mathcal N} =4$ superspace.

The superfields on the first D3-brane are functions of
$(x^0,x^1,\theta, \bar\theta;\, w,\bar w)$.  The superspace is
spanned by $(x^0,x^1,\theta, \bar\theta)$, while  $w = x^2 + ix^3$
should be thought of as a continuous index. The necessary $(2,2)$
superfields are a vector superfield $V$ and three chiral
superfields $\Phi$, $Q_1$, and $Q_2$.  While this resembles the
four-dimensional ${\mathcal N} = 1$ superfield content of the
${\mathcal N} =4$ theory, the component fields are distributed
very differently. The gauge connections $A_{0,1}$ are contained in
$V$, while $A_2 + i A_3$ is the lowest component of a chiral
superfield $\Phi$.  This chiral superfield transforms
inhomogeneously under U(N) gauge transformations with non-trivial
dependence on the index $w$. Gauge transformations are
parameterized by families of chiral superfields $\Lambda$ labelled
by $w,\bar w$. Two of the six adjoint scalars are contained in $V$
or, equivalently, the lowest component of a twisted chiral
superfield, which (in the abelian case) is $\Sigma = \bar D_+ \bar
D_- V$. The four remaining adjoint scalars comprise the lowest
components of the chiral superfields $Q_1$ and $Q_2$.  In $(2,2)$
superspace, the four-dimensional ${\mathcal N} =4$ action is
\begin{equation}
\begin{split}
S_{D3} &=  \frac1{g^2}\int d^2x\, d^2w\, d^4\theta \tr\Big(
\Sigma^\dagger\Sigma+(\d w+g\bar\Phi)e^{gV}(\d{\bar w} + g\Phi)e^{-gV}\\
&\qquad\qquad+ \sum_{i=1,2}e^{-gV}\bar Q_i e^{gV} Q_i\Big)\\
&\qquad +\int d^2x\, d^2w\, d^2\theta\tr\left(Q_1[\d{\bar
w}+g\Phi,Q_2]\right) + c.c.\,, \label{action}
\end{split}
\end{equation}
We shall take (\ref{action}) as the action for degrees of freedom
propagating on the first stack of D3-branes.  The superfields on
the second stack of D3-branes are functions of $(x^0,x^1,\theta,
\bar\theta;\, y,\bar y)$ where $y=x^4+ix^5$. Their contribution to
the action is just like (\ref{action}) with $N\rightarrow
N^{\prime}$, $w \rightarrow y$ and primes added to all
superfields.

The remaining degrees of freedom are strictly two-dimensional and
arise from strings stretched between the orthogonal stacks of D3-branes.
These are described by two chiral superfields $B$ and $\tilde B$
in the $(N, \bar N^{\prime})$ and $(\bar N, N^{\prime})$
representations of U(N)$\times$ U(N$^{\prime}$). Together these
form a $(4,4)$ hypermultiplet. The part of the action containing
these fields is
\begin{equation}\label{defct}
\begin{split}
S_{D3-D3'} &= \int d^2x d^4\theta\tr\left( e^{-gV'}\bar
Be^{gV}B+e^{-gV}\bar{\tilde B}e^{gV'}\tilde B\right)\\
&\qquad +\frac{ig}2\int d^2x\, d^2\theta\tr\left(B\tilde
BQ_1-\tilde BBQ_1'\right).
\end{split}
\end{equation}
\relax From now on, we will not write out the explicit dependence on the
coupling constant $g$ anymore,  which is easily reintroduced as it
always enters as a prefactor of the $V$ and $\Phi$ superfields.

The lowest components of $Q^1$ and $\Sigma$, as well as their
primed counterparts, correspond to adjoint scalars describing
fluctuations in the directions $x^{6,7,8,9}$ transverse to both
stacks of D3-branes.  This can be seen by noting that expectations
values for these fields give mass to the defect fields $B$ and
$\tilde B$.  The lowest component of $Q_2$  describes fluctuations
of the first stack of D3-branes in the $x^{4,5}$ directions which
are tangential to the other stack of D3-branes (the D3$^{\prime}$
branes). The lowest component of $Q_2^{\prime}$ describes
fluctuations of the the D3$^{\prime}$-branes in the directions
$x^{2,3}$ tangential to the D3-branes.

\subsection{Supersymmetric solutions: Branches of the moduli space}

To determine the supersymmetric vacua, we look for the  gauge
equivalence classes of solutions to the D- and F-flatness
equations. Earlier investigations of moduli spaces of theories
with impurities appeared in \cite{KS,BGK}. The vanishing of the
F-terms in this theory requires:
\begin{align}
F_{Q_1}&=\d{\bar w}q_2+[\phi,q_2]+\delta^{(2)}(w)b\tilde
b=0\label{fq1}\,,\\
F_{Q_2}&=\d{\bar w}q_1+[\phi,q_1]=0\label{fq2}\,,\\
F_\Phi&= [q_1,q_2]\label{fphi}\,,\\
F_{Q^{\prime}_1}&=\d{\bar y}q_2^{\prime}+[\phi^{\prime},q_2^{\prime}]
+\delta^{(2)}(y)\tilde b b=0\label{fq1p}\,,\\
F_{Q_2^{\prime}}&=\d{\bar
y}q_1^{\prime}+[\phi^{\prime},q_1^{\prime}]=0\label{fq2p}\,,\\
F_{\Phi^{\prime}}&= [q_1^{\prime},q_2^{\prime}]\label{fphip}\,,\\
F_B&=\tilde bq_1\delta^{(2)}(w)-q_1'\tilde b\delta^{(2)}(y)=0\label{fb}\,,\\
F_{\tilde B}&= q_1b\delta^{(2)}(w)-bq_1'\delta^{(2)}(y)=0
\label{fbt}\, .
\end{align}
Throughout this article, we write the scalars which are the lowest
components of a chiral superfield in lower-case.  The vanishing of
the D-terms requires
\begin{equation}
D= \d w\phi-\d{\bar
w}\phi^\dagger+[\phi,\phi^\dagger]+[q_1,q_1^\dagger]+[q_2,q_2^\dagger]
+ \delta^{(2)}(w)(bb^\dagger-\tilde b^\dagger\tilde b)=0
\label{Dterm} \, .
\end{equation}
In looking for solutions of these equations,  we shall take all gauge fields
to vanish; $\phi = \phi^{\prime}=0$.

First consider the D-term equation (\ref{Dterm}).
Assuming that the $q$-fields are regular
the coefficient of the $\delta$-function has to vanish. Thus we
find
\begin{equation}
\tilde b\tilde b^\dagger = b^\dagger b,\label{backandforth}
\end{equation}
a vev for $b$ always implies a similar vev for $\tilde b$.

We can simultaneously diagonalize $q_1$ and $q_1'$ at $w=0$ since they
are acted upon by different gauge groups.   (\ref{fbt}) then becomes
\begin{equation}
0=b_{i'j}q_{1jj}(0)-q_{1i'i'}'(0)b_{i'j}=b_{i'j}(q_{1jj}(0)- q_{1i'i'}'(0)),
\label{branches}
\end{equation}
where the indices $i,j$ and $i',j'$ denote $SU(N)$ and
$SU(N^{\prime})$ indices, respectively. (\ref{branches}) is
satisfied if  $b_{i'j}$ or $q_{1jj}(0)-q_{1i'i'}'(0)$ vanish. The
former corresponds to the Coulomb branch of the gauge theory where
the orthogonal branes are seperated,  while the latter corresponds
to the Higgs branch. A similar analysis holds for $\tilde b$
instead of $b$.  The fields $b$ and $\tilde b$ may only condense
on the Higgs branch, where they are massless.

One might worry that the $q$ field in the (\ref{branches}) is
always evaluated at $w=0$ as required by the $\delta$-function in
(\ref{fb}) and not at $w=q_1'$. Van Raamsdonk \cite{MvR} has
investigated this question from a T-dual perspective. He found
that in order to remedy this concern, (\ref{fb}) should contain
$\exp(\alpha' q_1'\partial_w)$ factors. However this exponential
is higher order in $\alpha'$ and can be neglected to the order we
are working at.

Equations (\ref{fphi}) and (\ref{fphip}) show that we can
simultaneously diagnolize $q_1,q_2$ and $q_1',q_2'$.   There is no
`non-commutative' or `fuzzy' geometry associated with this system
and we can treat all adjoint fields as effectively abelian.   The
brane recombination always occurs pairwise.  This means for
example that one can not simultaneously have non-zero $b_{11'}$
and $b_{12'}$,  which would recombine three branes. Pairwise
recombination is required because of the D-flatness constraint
$b^\dagger b = \tilde b \tilde b^{\dagger}$ and because equations
(\ref{fq1}) and (\ref{fq1p}) imply that $b\tilde b$ and $\tilde b
b$ are diagonal\footnote{This is different from the T-dual system
of D0-branes resolving as instantons in D4 branes: There, each
instanton sits in a $SU(2)$ subgroup and thus involves two D4
branes. It might however be that our system corresponds to the
limit of infinte separation of branes of dyonic instantons that
becomes regular after two T-dualities}.  Without loss of
generality,  we will from now on consider the abelian case, in
which one D3-brane intersects one D3$^{\prime}$ brane.

Equation (\ref{fq2}) implies that $q_1$ is
a holomorphic function of $w$, a condition on the embedding
coordinates that is well known to be
necessary for a supersymmetric brane configuration, see for example
\cite{KMG,BBH}.
The condition (\ref{fq1}) implies that
$q_2(w)$ is holomorphic except at the origin $w=0$.
\EPSFIGURE{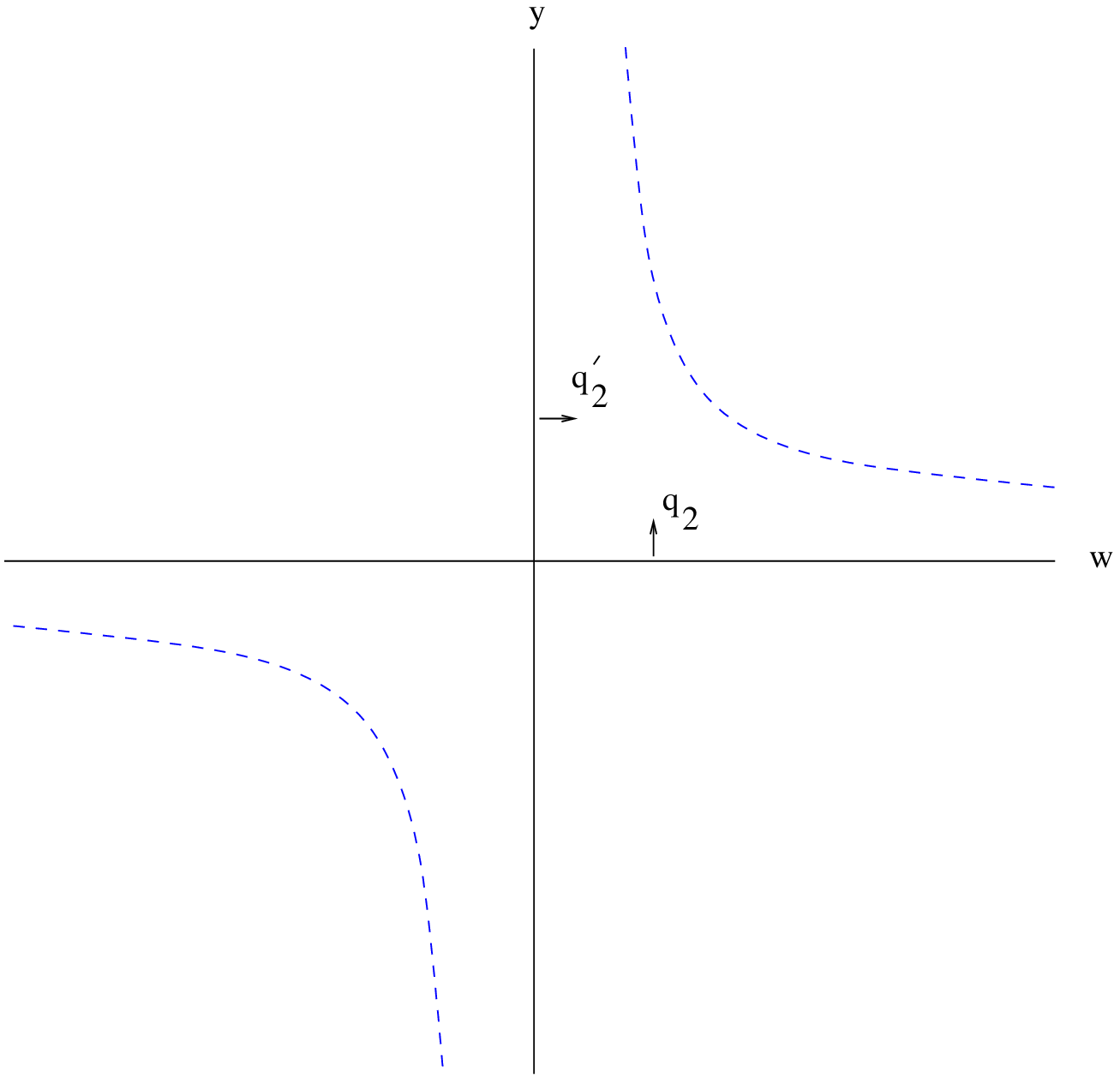,height=6cm}{Resolution of intersecting branes $wy=0$
to $wy =c$ on the Higgs
branch.}
\label{resolve}
The solution of  (\ref{fq1}) is
\begin{equation}\label{gnrl}
q_2(w) = \frac{b\tilde b}{2\pi i w} + h(w) \, ,
\end{equation}
where $h(w)$ is a holomorphic function of $w$.
It is easiest to study brane
recombination in the case in which $h(w)$ vanishes, corresponding to
boundary
conditions $q_2(w) \rightarrow 0$ at $w\rightarrow\infty$.
With boundary conditions at infinity corresponding to the original
configuration of orthogonal intersecting branes,
the unique solution of (\ref{fq1})
and
(\ref{fq1p}) is
\begin{equation}
q_2(w) = \frac{\tilde b b}{2\pi i w}\, ,  \qquad q_2^{\prime}(y) =
\frac{b\tilde b }{2\pi i y} \, . \label{solns}
\end{equation}
Recall that $q_2$ ($q_2^{\prime}$) describes the transverse fluctuations of
the D3-brane (D3$^{\prime}$-brane) in the
directions $y(w)$ tangential to the  D3$'$-brane (D3-brane).  The geometry
of the
D3-branes is
obtained by making the replacements $q_2 \rightarrow \alpha^{\prime}y$ and
$q_2' \rightarrow \alpha^{\prime}w$  in (\ref{solns}),  which can then be
collectively written as
\begin{align} wy = \frac{1}{2\pi i} b\tilde b \alpha^{\prime}.\end{align}
In other words the original pair of orthogonal branes now lie on
the same holomorphic curve.  Furthermore these branes have merged
such that there is actually only one D3-brane on the holomorphic
curve. This follows from the fact that the gauge group $U(1)
\times U(1)$ is broken to the diagonal $U(1)$ when $b\tilde b \ne
0$.

There are of course a much broader class of supersymmetric intersections and
holomorphic curves which
arise from brane recombination on the Higgs branch. This includes the
the recombination of branes at angles,
for which $h(y)$ is linear function of $y$\cite{CK}.

\section{Special Lagrangian three-folds and multiple brane intersections}

In the previous section, we discussed intersecting D-branes
preserving eight supersymmetries mer\-ging into a single
holomorphic curve on the Higgs branch.  We shall now consider
intersecting configurations which preserve four supercharges
and merge into a special Lagrangian three-fold on the Higgs
branch. Specifically,  we consider D6-branes  spanning four common
Minkowski space directions, while the remaining three directions
are embedded in a Calabi-Yau three-fold.

A special Lagrangian three-fold (see \cite{Joyce} for a review) is a
real three-dimensional surface embedded in a Calabi-Yau three-fold such that
\begin{align}
\omega|_L = {\rm Im}\, \Omega|_L = 0
\end{align}
where $\omega|_L$ is the restriction of the K\"ahler form to the surface
$L$,
while
$\Omega|_L$ is the restriction of the holomorphic three-form to $L$. Such
a manifold is calibrated with respect to ${\rm Re}(\Omega)$ and is volume
minimizing in its homology
class.  The role of the special Lagrangian conditions as a BPS
condition was first discussed in \cite{BBS}, (see also \cite{FF} and
references
therein).  However, the recombination of intersecting branes into
smooth special Lagrangian manifolds has not been explicitly discussed
from the point of view of the world-volume theory.

To study the geometries which arise on the Higgs branch of the
world-volume gauge theory of the intersecting D6-branes,  it is
most convenient to use a ${\mathcal N}=1$ superspace in four
dimensions, corresponding to the dimensions spanned by the
intersection. Note that our discussion is readily generalized to
intersecting branes of other dimension by T-duality.  For
intersecting D3-branes,  our discussion goes through almost
identically after replacing ${\mathcal N} =1, d=4$ superspace with
${\mathcal N}=2, d=1$ superspace (which is also associated with
four supercharges). We will begin by studying the action of a
stack of parallel D6-branes in a four-dimensional ${\mathcal N}
=1$ superspace\footnote{neglecting couplings to gravity/bulk
degrees of freedom}. It is then easy to construct the action for
intersecting configurations with four common directions and four
unbroken supersymmetries.

\subsection{7-dimensional maximally supersymmetric SYM in ${\mathcal N} =1,
D=4$ superspace}

The four dimensional ${\mathcal N} =1$ superspace representation
of the D6-brane action was constructed in \cite{AHGW} and
is related  to a construction discussed earlier in \cite{MSS}. The
four-dimensional ${\cal N} =1$ superfields entering the action
have the general form $F(x^{\mu},\theta,\bar\theta | \vec y)$,
where $(x^{\mu},\theta,\bar\theta)$ spans the four-dimensional
superspace, and $\vec y \sim (y^1,y^2,y^3)$ can be regarded as
continuous indices. The necessary degrees of freedom are contained
in three chiral fields $\Phi_I$ and a vector field $V$. The action
is
\begin{align}\label{actn}
S = & \frac{1}{g^2} \int d^3y\, d^4x\,d^2\theta\, \left. {\rm tr}
\left[ W_{\alpha}W^{\alpha} + \epsilon_{ijk}(\Phi_i
\frac{\partial}{\partial y^j} \Phi_k +
\frac{2}{3}i\Phi_i\Phi_j\Phi_k) \right]\right|_{\theta\theta} +
c.c.
\\ +  & \frac{1}{g^2} \left. \int d^3y\, d^4x\, d^4\theta\, {\rm tr}
\bar\Omega_i e^V \Omega_i e^{-V} \right|_{\theta\theta\bar\theta\bar\theta}
\, ,
\end{align}
where the indices $i,j,k$ take values from $1$ to $3$, and the superfield
\begin{align} \Omega_i = \Phi_i + e^{-V}(i\partial_i - \bar\Phi_i)e^{V} \, .
\end{align}
The scalar fields of the theory consist of seven gauge connections
$A_{0,1,2,3,4,5,6}$ and three Hermitian adjoint scalars
$X_{7,8,9}$ describing transverse fluctuations of the D6-brane.
These are distributed amongst the four-dimensional ${\mathcal N}
=1$ superfields as follows,
\begin{align}
V \rightarrow A_{0,1,2,3},  \qquad \Phi_i \rightarrow A_{4,5,6}\, ,
\;\; X^{7,8,9} \, .
\end{align}
The combination $A_{i+3} + iX^{i+6}$ for $i=1,2,3$ is the lowest
component of the chiral superfield $\Phi_i$.  Gauge
transformations are chiral superfields $\Lambda(\vec
x^{\mu},\theta; \vec y)$,  which act in the following way,
\begin{align}
e^{V} &\rightarrow
e^{i\Lambda^{\dagger}}e^{V}e^{-i\Lambda} \, ,\\
i\partial_i-\Phi_i &\rightarrow e^{i\Lambda}(i\partial_i -\Phi_i)
e^{-i\Lambda} \, .
\end{align}
Under these transformations,  the superfield $\Omega_i$,  transforms as
\begin{align}
\Omega_i \rightarrow e^{i\Lambda}\Omega_i e^{-i\Lambda}\, .
\end{align}
Seven-dimensional Lorentz invariance is not manifest but becomes
apparent upon integrating out auxiliary fields and performing
suitable field redefinitions. Note the resemblance of the
superpotential to a Chern-Simons action.

The action we have written is for a D6-brane in flat space,  in
which we have taken the SYM approximation of the Dirac-Born-Infeld
action. As we shall see shortly, a consideration of the full
Dirac-Born-Infeld action should lead to a modified D-flatness
condition but not a modified F-flatness condition. If the D6-brane
wraps a three-cycle of a Calabi-Yau, the D-terms will be further
modified, although the diffeomorphism invariant Chern-Simons
superpotential will not change.

\subsection{Special Lagrangians from F- and D-flatness}

Let us now consider the F- and D-flatness conditions for the action
(\ref{actn}).  Henceforward it is convenient to write
$X^{i+6}$ as $X^i$, $A_{i+3}$ as $A_i$ and
$\frac{\partial}{\partial y^i} = \partial_i$ where $i=1,2,3$.
F-flatness gives $\partial W/\partial \phi_i =0$, or
\begin{align}
&D_i X_j - D_j X_i =0 \, , \label{lagr}\\
&F_{ij} - [X_i, X_j] = 0 \, ,
\end{align}
where $F_{ij}$ is the gauge field strength. The D-flatness
condition is
\begin{align}
D_i X_i = 0 \, .
\end{align}
We consider only abelian solutions with $F_{ij} =0$, in which case
F- and D-flatness become
\begin{align}
&\partial_i X_j -\partial_j X_i = 0 \, ,\label{fterm}\\
&\partial_i X_i =0. \label{dterm} \end{align}

The solutions of (\ref{fterm}) and (\ref{dterm}) determine the
embedding of the D6-brane in $\mathbb{C}^3$,  which we shall
parameterize by the complex coordinates $u^i \equiv y^i +
i{\alpha'}X_i$.
It is now easy to show that equations (\ref{fterm}) and (\ref{dterm})
are linearized
special Lagrangian conditions\footnote{SLAG conditions were also discussed
in
terms of graphs of functions in \cite{Joyce} and in \cite{KLM},
although there in a different
formalism.}.  This can be seen as follows.
For $\mathbb{C}^3$, the K\"ahler form is $\omega = du^i
\wedge d\bar u^i$, while the holomorphic three-form is
$\Omega = du^1 \wedge du^2 \wedge du^3$.
Restricted to the embedding
$X_i(y^j)$, the differentials $du^i$ satisfy
\begin{align}
du^i = dy^i + i {\alpha'}\frac{\partial X_i}{\partial y^j}
dy^j \, .
\end{align}
So that the condition for a Lagrangian manifold is just
\begin{align}\label{Lagrangian}
\omega|_L = du^i \wedge d\bar u^i = 2i {\alpha'} dy^i \wedge dy^j \,
\partial_i X_j =0 \, ,
\end{align}
which is equivalent to F-flatness (\ref{fterm}).  A special
Lagrangian manifold satisfies the additional condition
\begin{align}
{\rm Im}\,\Omega|_L =&  {\rm Im} \, du^1 \wedge du^1 \wedge du^3
\nonumber
\\ =& \, dy^i \wedge dy^j \wedge dy^k \,
{\alpha'}\left(\partial_i X_i + {\alpha'}^2{\rm det}
(\partial_i X_j)\right) = 0. \label{special}
\end{align}
The determinant in (\ref{special}) is with respect to the matrix
indices $ij$. If we use (\ref{fterm}), to write $\vec X$ as a
gradient of a scalar potential $f$, then we recognize in
\ref{special} the three dimensional version of the special
Lagrangian condition $0=\det(Id+{\rm Hess}(f))$ that was discussed
in \cite{Joyce}. Up to the determinant term,  (\ref{special})
agrees with the D-flatness condition (\ref{dterm}).  We assume
that the determinant arises from the D-flatness condition for the
full Dirac-Born-Infeld Lagrangian.  Unfortunately, the
supersymmetric versions of Dirac-Born-Infeld actions are quite
involved and the known superspace descriptions \cite{Ketov} do not
include scalar fields.

In the special case in which ${\rm det} (\partial_i X_j)$
vanishes, the solutions of the F- and D-flatness equations are
shared by the SYM and DBI actions. This is similar to the case of
the BIon for which certain solutions are shared among Yang-Mills
and Born-Infeld theories as the non-linear terms of the
Born-Infeld field equation vanish for them \cite{Gibbons,
Townsend}. We  find such a situation for the Higgs branch of a
triple D6-brane intersection.  The special Lagrangians which arise
in this case are the product of a holomorphic curve with a line,
for which the determinant vanishes.

Note that if one were to view $X_i$ as a gauge field,  (\ref{Lagrangian})
and
(\ref{special}) would require this gauge field to be a flat connection in
a particular non-linear gauge. This is a very special gauge condition in the
following sense.  The $SU(3)$ invariance of the conditions
(\ref{Lagrangian}) and (\ref{special}) means that one can exchange
coordinates
with ``gauge'' fields such that connections remain flat and the form of
the gauge condition is unchanged.

\subsection{Knots, non-compact three-cycles, and resolved SLAG
intersections}

The F- and D-flatness equations,
\begin{align}
&\partial_i X_j - \partial_j X_i = 0 \, , \label{Ft2}\\
&\partial_i X_i + {\alpha'}^2 {\rm det} (\partial_i X_j) =
0\, , \label{Dt2}
\end{align}
have solutions which can be interpreted as flat connections
subject to a particular gauge condition. From
the example of the double intersection discussed in section 2, we
expect
that it is also necessary to include delta function sources in these
equations
when considering the recombination of intersecting branes.
In particular, the flat connection condition is overly restrictive, as can
be
seen in the following example.

Consider a special Lagrangian three-cycle ${\cal C} \times \mathbb{R}$
where ${\cal C}$ is the holomorphic curve $w=c/v$, with $c$ real
and
\begin{align}
w \equiv X_2 + i X_1\, , \qquad v = y^1 + i y^2 \qquad X_3 = 0 \, .
\end{align}
Since $\partial_{\bar v} w = 2\pi i c \delta^2(v)$, it follows
that
\begin{align} &\partial_1 X_2 - \partial_2 X_1 = 2\pi c
\, \delta^2(y^1,y^2)\, , \qquad \partial_2 X_3 - \partial_3 X_2 =
\partial_1 X_3 - \partial_3 X_1 = 0\, ,  \label{Ft3}\\
&\partial_1 X_1 + \partial_2 X_2 = 0. \label{Dt3}\end{align}
\relax From the point of view of the D6-brane world-volume theory, the first
set of conditions (\ref{Ft3}) arises from a modification of the
(abelian) Chern-Simons superpotential which includes a holomorphic
Wilson line:
\begin{align}
W = \int_{\Sigma_3}{ \phi d \phi} + i 2\pi c \int_{y^1 =y^2 =0}
dy^3\phi_3 \, ,
\end{align}
while the D-term is unchanged.  In the limit $c\rightarrow 0$ one obtains an
intersection of complex planes $wv=0$ times a line.

One can also include a more general
holomorphic Wilson line in the superpotential
\begin{align}\label{supp}
W = \int_{\Sigma_3}{ \phi d \phi} + i c 2\pi \int_{\Sigma_1}\phi
\, , \end{align}
where $\Sigma_1$ is an arbitrary closed or infinite path $\vec
y^i(s)$. The real part of the F-term is not modified by the Wilson
line, such that F-flatness still requires a flat gauge connection.
However the imaginary part of the F-term is corrected by a term with delta
function support on $\Sigma_1$;
\begin{align} \label{mod}{\rm Im} F_{\Phi_k} = \epsilon_{ijk}(\partial_i X_j
-
\partial_j X_i) - J_k = 0,\end{align}
where
\begin{align}
J_k =  \int ds \delta^{3}(\vec y - \vec
y(s))\frac{dy^k(s)}{ds}. \end{align}
The solutions of (\ref{mod}) belong
to a class of non-compact Lagrangian
manifolds\footnote{Despite the delta function the Lagrangian
condition $\omega|_{\cal L} =0$ holds at any finite point on the
(non-compact) curve which is a solution of the F-flatness
equation.} associated with the path $\Sigma_1$.

In the context of the deformed conifold $T^*(S^3)$, the existence
of a Lagrangian manifold passing through every knot in $S^3$  was
pointed out in \cite{OV}.  With a different motivation in mind,
the authors of \cite{OV} considered topological open strings in
the background of a D6-brane wrapping $\Sigma_3 = S^3$ and a
D6-brane on a non-compact Lagrangian manifold intersecting $S^3$
over a knot $\Sigma_1$. The solutions of (\ref{mod}) in this
setting correspond to a recombination of these D6-branes into a
single D6-brane on a smooth Lagrangian manifold. For topological
strings one does not consider the D-flatness condition.

For non-topological strings,  one must also consider D-flatness, which
leads to a special Lagrangian manifold.
Although the D-term only takes the
form (\ref{Dt3}) on $\mathbb{C}^3$, we expect it still plays a role
analogous to a gauge fixing for a non-trivial Calabi-Yau manifold.

For the case of $\mathbb{C}^3$,  solutions of (\ref{mod}) and
(\ref{Dt3}) are smooth non-compact special Lagrangian manifolds
which should reduce to an intersection in the limit $c\rightarrow
0$.  As noted above D-flatness can be regarded as a gauge fixing.
There may be interesting non-perturbative effects such as Gribov
ambiguities.  However a unique perturbative expansion can be found
if the path $y^i(s)$ is almost straight over distances of order
$\sqrt{\alpha^{\prime}}$ and suitable boundary conditions are
imposed at infinity. The perturbative expansion in
$\alpha^{\prime}$ is obtained by writing
\begin{align}
\vec X = \sum_{n=0}\vec X_{(n)}{\alpha'}^2
\end{align}
and solving
\begin{align}
\vec\nabla\times\vec X_{(0)} &= \vec J, \qquad \vec\nabla\cdot\vec
X_{(0)}
= 0 \, , \nonumber \\
\vec\nabla\times\vec X_{(n>0)} &= 0, \qquad \vec\nabla\cdot\vec
X_{(n>0)} =
{\rm det}(\vec\nabla\otimes\vec X_{(n-1)}).
\end{align}

The physical origin of the Wilson line in the effective
superpotential (\ref{supp}) is the condensation of degrees of
freedom localized at the intersection of the brane we have been discussing
and
another brane which meets it over the common superspace directions
and the path $\Sigma_1: y(s)$.  Since the
coordinate $s$ is transverse
to the superspace, kinetic terms in this direction should arise from a
superpotential. For the degrees of freedom at the intersection, this
superpotential takes the form
\begin{align}
W_{\rm impurity} = \int ds B(s)\left(\partial_s - i\left(\Phi_i(\vec
y(s)) - \Phi_i^{\prime}(\vec y(s))\right) \frac{dx^i}{ds}\right) \tilde
B(s) \, ,
\end{align}
analogous to (\ref{defct}). Here $B(s)$ and $\tilde B(s)$ are an
infinite class of chiral superfields labelled by $s$ which
describe the degrees of freedom at the intersection.  $B$ is in
the bifundamental representation of the gauge group on the two
D6-branes,  while $\tilde B$ is in the conjugate representation.
The chiral superfield $\Phi_i^{\prime}$ is the counterpart of
$\Phi_i$ on the second D6-brane.  A similar expression also
appeared in \cite{OV}.  Note that integrating out F-terms for $B$
and $\tilde B$ gives bosonic kinetic terms in the $s$ direction.
The effective superpotential (\ref{supp}) arises if the impurity
fields $B$ and $\tilde B$ condense,  i.e. if $\langle b\tilde b
\rangle =c$ where $b,\tilde b$ are the lowest components of
$B,\tilde B$. F-flatness now gives
\begin{align}\label{ftrm}
F_{\Phi_i} = 0= \epsilon_{ijk}(\partial_j \phi_k - \partial_k \phi_j) -
\int ds  \delta^3(\vec y -\vec y(s))\frac{dy^i(s)}{ds} \tilde b(s)
b(s)\, .
\end{align}
Note that $\partial_i F_{\Phi_i} =0$ implies $\partial_s b(s)\tilde b(s) =
0$.
Recalling that $\phi_i = A_i +i X_i$, solutions of (\ref{ftrm})
with vanishing gauge fields
$A_i$ require $b\tilde b$ to be real.

There is also a K\"ahler potential for the degrees of freedom at the
intersection, which, at least in the $\alpha^{\prime} \rightarrow
0$ limit, is of the form
\begin{align}
\int ds \sqrt{g_{ss}}(\bar B(s) e^{V(\vec y(s))} B(s)e^{-V'(\vec y(s))} +
\bar
{\tilde B}(s) e^{-V(\vec y(s))} \tilde B(s)e^{V'(\vec y(s))}). \end{align}
The
D-flatness condition is modified by the K\"ahler term to become
\begin{align}
i\partial_i X_i + \int ds \sqrt{g_{ss}}\delta^3(\vec y - \vec
y(s))(b^* b - {\tilde b}^* \tilde b) = 0 \, .
\end{align}
The new condition arising from this is the real part $b^* b -
{\tilde b}^* \tilde b = 0$,  which we assume is unaltered when
considering the full Dirac-Born-Infeld action.

The vanishing of the F-terms for $\tilde B$ and $B$ requires
\begin{align}\label{carp}
F_{\tilde B} &= \left(\partial_s -
i \left(\phi_i(\vec y(s))-\phi'_i(\vec y(s))\right)
\frac{dy^i(s)}{ds}\right)\tilde b \nonumber\\ & =\left(\partial_s +
i \left(\phi_i(\vec y(s))-\phi'_i(\vec y(s))\right)
\frac{dy^i(s)}{ds}\right) b =0\, . \end{align}
A similar equation applies for $\tilde b$.  Since $b\tilde b$ is real and
independant of $s$, and $|b| = |\tilde b|$, we can write
$b = \sqrt{c}\exp(i\theta(s)), \tilde b = \sqrt{c}\exp(-i\theta(s))$.
For solutions with vanishing gauge field, the real and imaginary parts of
(\ref{carp}) imply that $\partial_s(\theta(s))=0$ and
\begin{align}
\left(X_i(\vec y(s)) - X_i^{\prime}(\vec
y(s))\right)\frac{dy^i(s)}{ds}\sqrt{c} = 0.
\end{align}
This equation distinguishes between the Coulomb branch for which $c=0$ and
the
branes can separate, and the
Higgs branch for which
\begin{align}
\left(X_i(\vec y(s)) - X_i^{\prime}(\vec
y(s))\right)\frac{dy^i(s)}{ds} = 0.\end{align}
On the Higgs branch, the solutions of the F- and D-flatness
conditions correspond to the recombination of the intersecting
D6-branes into a smooth special Lagrangian manifold.  Note that
the formalism we have been using generalizes trivially to
intersections of branes of other dimensions which also preserve
four supercharges.  The  one-dimensional ${\mathcal N} =2$
superspace action for D3-branes is just a dimensional reduction of
(\ref{action}) in which $d^4 x \rightarrow dt$.

\subsection{Triple and quadruple intersections of branes preserving four
supercharges}

We now consider a configuration of D6-branes in flat space which
intersect over three common spatial directions and preserve four
supersymmetries.  Writing the D6-brane action in a
four-dimensional ${\mathcal N} =1$ superspace as in (\ref{actn})
greatly facilitates the construction of the action for this type
of intersection.   Once we have constructed the action for the
intersection we  examine the geometry which arises on the
Higgs branch.  In this case,  the special Lagrangians which we
obtain are holomorphic curves times a line.

The D6-brane orientations which we consider are summarized in the
following table.
\newpage

\TABLE{
\begin{tabular}{|c|cccccccccc|}
\hline
&0&1&2&3&4&5&6&7&8&9\\
&$z_0$&$z_1$&$z_2$&$z_3$&$y_{14}$&$y_{24}$&$y_{34}$&$y_{23}$&$y_{31}$&$y_{12}$\\
\hline
1&\ $\x$&\ $\x$&\ $\x$&\ $\x$& \ $\x$&&&&\ $\x$&\ $\x$\\
2&\ $\x$&\ $\x$&\ $\x$&\ $\x$& &\ $\x$&&\ $\x$&&\ $\x$\\
3&\ $\x$&\ $\x$&\ $\x$&\ $\x$& &&\ $\x$&\ $\x$&\ $\x$&\\
4&\ $\x$&\ $\x$&\ $\x$&\ $\x$&\ $\x$&\ $\x$&\ $\x$&&&\\
\hline
\end{tabular}
}
\label{tabl1}

\noindent The $\x$'s indicate a direction in which the D6-brane is
extended. If there are D6-branes in three or all four of these
orientations, then four supersymmetries are unbroken.  We shall
focus on the case in which there are D6-branes in the
first three orientations, in which case it will still
be convenient to use the above notation.
The directions $z^{0,1,2,3}$ belong to the four dimensional superspace.
It is convenient to label the six
coordinates transverse to the superspace by $x_{AB}$ where $A\ne
B$, $A$ and $B$ run from $1$ to $4$,  and there is no distinction
between $y_{AB}$ and $y_{BA}$.  The $A$th stack of branes then
extends in the directions $y_{AB}$ for three values of $B$.
For example stack two
extends in $y_{12}, y_{23}$ and  $y_{24}$ while it is localized
in $y_{13}$, $y_{14}$, and $y_{34}$.

Associated to the D6-brane of the A'th orientation is a vector
multiplet $V_A$ and three chiral multiplets  $\Phi^A{}_B$ where
$B\ne A$.  
The lowest component of $\Phi^A{}_B$ contains the gauge connection
in the $y_{AB}$ direction and the scalar that describes the
position in direction $y_{CD}$,  where none of the labels $A,B,C$
or $D$ are equal.  Note that this is also the one direction in which
neither brane A nor brane B are extended.

In this notation, the action for a D6-brane in
the A'th orientation is
\begin{equation}
\begin{split}
&S_A=\frac{1}{g^2}\int\, d^4 z\, d^4\theta \,\prod_E
dy_{AE}\,\,\sum_{B\ne A} \tr\left( e^{-V_A} \bar\Omega^A{}_B
e^{V_A}
\Omega^A{}_B\right) \\
&+\frac1{g^2}\int\, d^4 z\, d^2\theta\,\prod_E
dy_{AE}\,\,\sum_{BCD}\epsilon_{ABCD} \tr \left( \Phi^A{}_B
\partial_{AC} \Phi^A{}_D+i\frac 23 \Phi^A{}_B
\Phi^A{}_C \Phi^A{}_D\right)\\
&+c.c. \, ,
\end{split}
\end{equation}
where
\begin{align} \Omega^A{}_B \equiv \Phi^A{}_B +
e^{-V_A}(i\partial_{AB} - \bar \Phi^A{}_B)e^{V_A}\, .
\end{align}
The action is invariant under the gauge transformations
\begin{equation}
\begin{split}
(i\partial_{AB}-\Phi^A{}_B)&\longrightarrow
e^{i\Lambda_A}(i\partial_{AB}-\Phi^A{}_B)e^{-i\Lambda_A}\, ,\\
e^{V_A}&\longrightarrow \,
e^{i\Lambda_A^\dagger}e^{V_A}e^{-i\Lambda_A} \, .
\end{split}
\end{equation}

There are also additional fields that live at the intersection of
pairs of branes with different orientations. These degrees of
freedom are chiral superfields $B_{AB}$, where $B_{AB}$ and
$B_{BA}$ are not equivalent. Under gauge transformations
\begin{align}
B_{AB} \rightarrow e^{i\Lambda_A}B_{AB} e^{-i\Lambda_B}, \qquad
B_{BA} \rightarrow e^{i\Lambda_B} B_{BA} e^{-i\Lambda_A} \, .
\end{align}
The fields $B_{AB}$ depend on the parameter $y_{AB}$ in addition
to superspace coordinates shared by all the branes. The action
contains the terms
\begin{equation}
\begin{split}
&S_{AB}= \frac1{g^2}\int\!\! d^4 z \,d^4\theta \, dy_{AB}\;\;
  \tr\, \left( e^{-V_B}\bar B_{AB}e^{V_A}B_{AB}+e^{-V_A}\bar{
  B}_{BA} e^{V_B} B_{BA}\right) \\
&+\frac1{g^2}\int\!\! d^4 z\, d^2\theta \, dy_{AB} \;\;
  \tr\, \left(B_{BA}\partial_{AB} B_{AB} + iB_{BA}\Phi^A{}_B
B_{AB}-i\Phi^B{}_A
   B_{BA} B_{AB}\right) + c.c.
\end{split}
\end{equation}
The kinetic terms in the $y_{AB}$ direction come from the the
superpotential rather than the K\"ahler potential.

Note that the term $S_A + S_B + S_{AB}$ is fixed by the
requirement that it be the action for a double intersection which
preserves eight supercharges. Upon compactifying $z^2, z^3$ and
$y_{AB}$, one must have the action of the double D3-intersection
discussed in section 2 (after integrating out auxiliary fields and
performing some field redefinitions). Since we consider D6-branes
in three or four of the orientations in table (\ref{tabl1}) there
may also be a term in the superpotential involving the twist
fields of more than one pair of branes. This term is not fixed by
any symmetry,  and must be obtained from a string scattering
calculation. Since three or more D6-branes intersect over four
dimensions, this term is defined only on the superspace
coordinates.  For the triple intersection, the general form of
such a term is
\begin{equation}\label{locsup}
S_{ABC}=\frac1{g^2}\sum_n \, \gamma_{n} \, \int\, d^4z\, d^2\theta\,  B^n
\end{equation}
where $B^n$ is shorthand for a gauge invariant product of $n$ twist fields
in which  all three brane indices are represented. As in the case of
D3-branes, the $B$'s are dimensionless\footnote{The
dimension of $g$ is $[g]=-3/2$ as appropriate for the Yang-Mills
coupling in seven dimensions.}.
Thus $\gamma_n$ are numbers
which must be determined from a  string scattering computation.
The simplest possible term in (\ref{locsup}) is $B_{AB}B_{BC}B_{CA}$
The coefficient $\gamma_3$.
would most easily read of as the strength of a Yukawa coupling
between three twist fields. This calculation has been performed in
\cite{Abel,CP,CIM} and $\gamma_3$ is implicit in those results. As
long as it does not vanish, the precise numerical value of
$\gamma_3$ is not important for our purposes.
For the quadruple action there may also be terms $S_{ABCD}$
in which all four brane-indices are represented.
The complete action for the system of intersecting branes is the
sum of all these single, double, triple and possibly quadruple
intersection actions:
\begin{equation}
S=\sum_A S_A+\sum_{AB} S_{AB}+\sum_{ABC} S_{ABC}+ S_{ABCD} \, .
\end{equation}

\section{F- and D-flatness for the triple intersection}

We now look for the solutions of the F- and D-flatness equations to
see what geometrical configurations arises when the twist fields
condense.  Let us first consider the vanishing of the F-term for
$\Phi^A_B$:
\begin{equation}
  0=F_{\Phi^A{}_B}=\sum_{CD}\epsilon^A_{BCD}( \partial_{AC}\phi^A{}_D+
  i\phi^A{}_C\phi^A{}_D) + \left.i\delta^2(y_{AE})\right|_{E\ne B}b_{AB}
b_{BA} \, .
  \label{FPhi}
\end{equation}
As before, we are interested only in solutions in which the gauge
connections vanish. Thus all the $\phi$'s are anti-hermitian,  and
we shall write $\Phi^A_B = i X^A_B$.  Considering the Hermitian
part of (\ref{FPhi}) gives
\begin{align}
\sum_{CD}\epsilon^A_{BCD}[X^A_C, X^A_D] =
\left.\delta^2(y_{AE})\right|_{E\ne B} (b_{AB}b_{BA} -
b_{BA}^{\dagger}b_{AB}^\dagger) \, . \label{fphiAB}
\end{align}
As long as the $X's$ are regular,  the left and right hand side of
(\ref{fphiAB}) must vanish separately.  This precludes a
non-commutative geometry, and we will henceforward just consider
abelian equations. The anti-Hermitian part of (\ref{FPhi}) gives
\begin{align}\label{curr}
\sum_{CD}\epsilon^A_{BCD}\partial_{AC}X^A_{D} = (b_{AB}b_{BA} +
b_{BA}^{\dagger}b_{AB}^\dagger)\delta^2(y_{AE})|_{E\ne B}\, .
\end{align}
which is a special case of (\ref{mod}).  Regarding $X$ as a
magnetic field, equation (\ref{curr}) can be viewed as a
magnetostatics equation $\vec\nabla\times\vec B = \vec J$.  For
the case of the triple intersection,  there are two orthogonal
lines of current as the index $B$ in (\ref{curr}) can take two
possible values.  Current conservation then requires that
$\partial_{AB}(b_{AB}b_{BA})$ vanishes (where there is no sum on
any of the indices). We therefore look for solutions for which
$b_{AB}b_{BA}$ is constant.

Next consider the vanishing of the D-term, which in the abelian
case is:
\begin{equation}
\begin{split}
0&=D_{V_A}\\
    &=\sum_{B\ne A} \partial_{AB}(\phi^A{}_B - \bar\phi^A{}_B)  +
    \sum_{BCD}\epsilon_{ABCD}
\delta(y_{AC})\delta(y_{AD})\left(b_{AB}b_{AB}^{\dagger}-
    b_{BA}^\dagger b_{BA}\right) \, .
\end{split}
\label{DV}
\end{equation}
Considering the real and imaginary parts of (\ref{DV}) separately
gives
\begin{align}
b_{AB}b_{AB}^{\dagger}-b_{BA}^\dagger b_{BA} = 0
\end{align}
and
\begin{align}\label{dcons}
\sum_{B}  \partial_{AB}{X^A_B} =0\, ,
\end{align}
which corresponds to (\ref{Dt2}) if ${\rm
det}_{BC}\partial_{AB}X^A_C =0$.

Finally consider the F-term for the twist fields:
\begin{equation}
  0=F_{B_{AB}}=\partial_{AB} b_{BA}+i\phi^A{}_B
  b_{BA}-i b_{BA}\phi^B{}_A \, + \, \delta(y_{AB}) \frac{\partial
  W_L}{\partial b_{AB}}
\, .\label{FBAB}
\end{equation}
In light of the previous F- and D-flatness conditions,  we can write
$b_{AB} = c_{AB}e^{i\theta_{AB}(y_{AB})}$ and $b_{BA} =
c_{AB}e^{-i\theta_{AB}(y_{AB})}$.  Taking $\phi^A_B = iX^A_B$, (\ref{FBAB})
becomes
\begin{align}
ic_{AB} \partial_{AB}\theta_{AB}(y_{AB}) -  (X^A_B - X^B_A)c_{AB}
+ e^{-i\theta_{AB}(y_{AB})} \frac{\partial W_L}{\partial
b_{AB}}\delta(y_{AB}) =0 \, ,
\end{align}
where $W_L$ is the part of the superpotential localized at the point which
where all the branes intersect, $S_{ABC} \sim \int W_L$.
Solutions for which X is regular
(describes a smooth surface) and
$\partial_{AB} \theta_{AB} = 0$ require
\begin{align}
\frac{\partial W_L}{\partial b_{AB}} = 0\label{localsup} \, , \\
(X^A_B - X^B_A)c_{AB} = 0 \label{coulombhiggs} \, .
\end{align}
Solutions of equation (\ref{coulombhiggs}) distinguish between the
Coulomb branch with $c_{AB} =0$ and the Higgs branch with $X^A_B-
X^B_A =0$. If one had only the lowest order term, $W_L \sim \gamma_3
B_{AB}B_{BC}B_{CD}$, (\ref{localsup}) would require $b_{AB}$ to
vanish for all but one pair of indices $A$ and $B$. We have no
reason to exclude higher order terms in $W_L$, however it is easy
to see that a vanishing $b_{AB}$ for all but one pair of indices
$A$ and $B$ remains a solution of (\ref{localsup}). This is
because $W_L$ must involve all three indices, since terms with
just one pair of indices would live in five rather than four
dimensions. Actually, gauge invariance requires each index to appear
at least twice.
New solutions will in general exist in the presence of higher order
terms. However, for a canonical normalization of the $B$ fields,
such that the two point function is independent of $g$ in the weak
coupling limit, the extra solutions approach infinity in field
space in the weak coupling limit.  We will not consider such
solutions here.

When the $b_{AB}$ condense for one pairing of indices, two branes
recombine on the Higgs branch while the third is not deformed. The
same configuration was also obtained by Lambert \cite{Lambert}
using a supersymmetry condition arising from bulk supergravity.
For the branes which recombine, (\ref{dcons}) and (\ref{curr})
have the structure $\vec\nabla\times\vec X(\vec y)=
c^2\delta^2(y^1,y^2)$ and $\vec\nabla\cdot\vec X(\vec y) = 0$,
whose solution\footnote{This solution is unique if one imposes
boundary conditions such that calibrated surface is the
asymptotically the same as the original intersection.} is
\begin{align}
X^3 = 0, \qquad X^2 + i X^1 = \frac{c^2}{y^1 + i y^2}\, .
\end{align}
This corresponds to a holomorphic curve $\Sigma$ constrained to lie in
the $X^3=0$ plane, times a line parametrized by the coordinate $y_3$.
Furthermore the third D6 brane remains flat.
To be more specific,  consider the case in which $b_{12}$ and $b_{21}$
condense.  We then have the following solution for each of the three
orthogonal branes:
\begin{align}
X^1_2 &=0, \qquad X^1_3 + i X^1_4 = \frac{c}{y_{14} +iy_{31}}\, ,
\label{oner} \\
X^2_1 &=0, \qquad X^2_3+ i X^2_4 = \frac{c}{y_{24} + iy_{23}}\, ,
\label{twor}\\
X^3_1 &= X^3_2 = X^3_4 =0 \label{threer}.
\end{align}
The $X$ fields are identified with fluctuations in a particular direction
as
\begin{align}
\begin{split}
X^1_2 \sim y_{34},\qquad  X^1_3 \sim y_{24},\qquad X^1_4 \sim y_{23} \,,\\
X^2_1 \sim y_{34},\qquad X^2_3 \sim y_{14},\qquad X^2_4 \sim y_{31} \,,\\
X^3_1 \sim y_{24},\qquad X^3_2 \sim y_{14},\qquad X^3_4 \sim
y_{12}\, .\end{split} \end{align}
Thus (\ref{oner}) and (\ref{twor})
describe a holomorphic curve times a line, while (\ref{threer}) is
a special Lagrangian plane. Altogether, we may write the special
Lagrangian manifold $L$ as
\begin{gather}
L = \left\{ y_{34} =0, \;\;
(y_{24}+ i y_{23})(y_{14} + i y_{31}) =c\, \right\}  \cup
\left\{y_{14}=y_{24}=y_{12}=0\right\}\, ,
\end{gather}
which corresponds to
\begin{gather}
L \simeq \Sigma \times \RR \cup \RR^3 \label{LL}
\end{gather}
It is instructive to compare this result with the analysis of
Gukov and Tong \cite{GT}, who use D6 brane intersections in order
to construct manifolds of $G_2$ holonomy in M-theory. For a
configuration of three D6 branes  intersecting at angles of
$2\pi/3$ they find a very similar structure to (\ref{LL}). Here we
obtain this structure from F- and D-flatness in the world-volume
gauge theory.


The Higgs branch of the triple intersection provides a controlled
setting to illustrate the Higgs mechanism which is expected to occur
in intersecting brane constructions containing the Standard Model
\cite{BGKL1, BGKL2, AFIRU}. Typically, there one has (besides
the brane that yields the hyper-charge) three stacks of branes whose
gauge groups one identifies with the left handed $SU(2)$ and right
handed $U(1)$ and the color $SU(3)$. The fields at the intersections
of the two electro-weak stacks are the Higgs fields whereas the quarks
live at the other two intersections as they are in the bifundamental
of the color and the electroweak symmetry group.  In this context the
Higgs is tachyonic.  Unlike the supersymmetric brane recombination we
have described, a precise treatment of brane recombination in the
intersecting brane-world models is lacking and would seem to require
string field theory, although some interesting effective field theory
descriptions have been proposed \cite{Hashimoto,HT,Huang,Sato}.  Upon
condensation of the
Higgs field, the two ``electro-weak'' branes resolve into one giving a
diagonal subgroup after chiral symmetry breaking. The stacks
corresponding to the left- and right-handed weak gauge groups do not
intersect anymore with the stack of the color gauge group (as we found
above: we can only give one $B$ a vev as this renders the others
massive), making the quarks that stretch between the color and the
weak stacks massive.

\section{Scattering between recombined branes}

We now change gears somewhat and, returning to the case of a
D3-brane on the holomorphic curve $wy =c$, compute the
transmission amplitude for a particle at $x\rightarrow\infty$ to
reach $y\rightarrow \infty$.

This is similar to a calculation performed in \cite{CM} for the
BIon \cite{Gibbons}.  We will find that the transmission
approaches $1$ for $c\rightarrow\infty$ and vanishes for
$c\rightarrow 0$. This result is consistent with the expectation
that the D3-brane on the holomorphic curve becomes two separate
orthogonal branes at the point $c=0$.

To compute the transmission amplitude we will study the two-point
function or propagator of light fields living on the resolved
brane intersection.  Consider the massless wave equation with
respect of the induced metric on the brane. The 2-point function
will obey this covariant wave equation:
\begin{equation}
0=\BOX_g\psi
\end{equation}
where $\BOX_g$ is the d'Alambertian corresponding to the metric
that is pulled back to the curve $wy=c$. After a rotation in the
$w$ and $y$ planes we can assume that $c$ is real and
non-negative. We write $y=u+iv$ and $w=r\, e^{i\phi}$ and find
\begin{equation}
\begin{split}
ds^2&= dr^2+r^2d\phi^2+du^2+dv^2\\
&= \left(1+\frac{c^2}{r^4}\right)(dr^2+r^2d\phi^2).
\end{split}
\end{equation}
We see that this metric has the expected asymptotics: For large $r$
the prefactor trivializes and we have a flat two-dimensional
metric. For $r\to 0$, the prefactor becomes $c^2/r^4$ and after a
change of variables $u=\sqrt{c}/r$ again we have a flat
metric.

The two-point function from $r=\infty$ to $r=0$ tells us how much
of a wave that comes from infinity on the horizontal branch of the
brane is transmitted to the vertical branch of the brane. This
problem is like a quantum mechanical scattering problem,
associated with which is a concerved current
\begin{equation}
j=\bar\psi d \psi - \psi d\bar\psi,\qquad d*j=0.
\end{equation}

Unfortunately, we do not have an analytic solution to the wave
equation in the above metric. So we have to resort to numerical
methods. However, in the asymptotic regions $r\to\infty$ and $r\to
0$ the radial wave equation in flat space can be solved in terms
of Hankel functions $H_\pm$. For concreteness we consider the
S-wave for which $\psi$ is only a function of $r$. The case of
non-trivial $\phi$-dependence of can be treated similarly.

At large $r$, every solution to the wave equation has the form
\begin{equation}
\psi(r) = A H_+(r)+BH_-(r),\qquad H_\pm(r)\to\sqrt{\frac1r}e^{\pm
ir}.
\end{equation}
$A$ is the coefficient for the outgoing wave, $B$ correspondingly for
the ingoing one. Accordingly the current is
\begin{equation}
j_r=2riH_+(r)H_-(r)(|A|^2-|B|^2)dr.
\end{equation}
For $r\to 0$ we have similarly
\begin{equation}
\psi(r) = a h_+(r)+bh_-(r),\qquad h_\pm(r)=H_\pm\left(\frac{c}
{r}\right)
\label{asympsol}
\end{equation}
and
\begin{equation}
j=-\frac{ic}{r}h_+(r)h_-(r)(|a|^2-|b|^2)dr.
\end{equation}
Now we start with initial values for large $r$ that correspond to
$B=0$ and integrate the full equation numerically using {\it
mathematica}. At small $r$ we fit this numeric solution to the
form (\ref{asympsol}) and read of the coefficients $a$ and $b$. By
equating the two asymptotic expressions for the current $j_r$,
which satisfies $\partial_r j_r =0$ we find
\begin{equation}
1=\frac{r^2H_+(r)H_-(r)}{ c h_+(r)h_-(r)}\, \frac{|A|^2}{
|a|^2}+\frac{|b|^2}{ |a|^2}.
\end{equation}
The first term is the coefficient of transmission, the second one the
coefficient of reflection. Figure 3 shows the transmission plotted
against the value of $\tilde bb$:
\EPSFIGURE{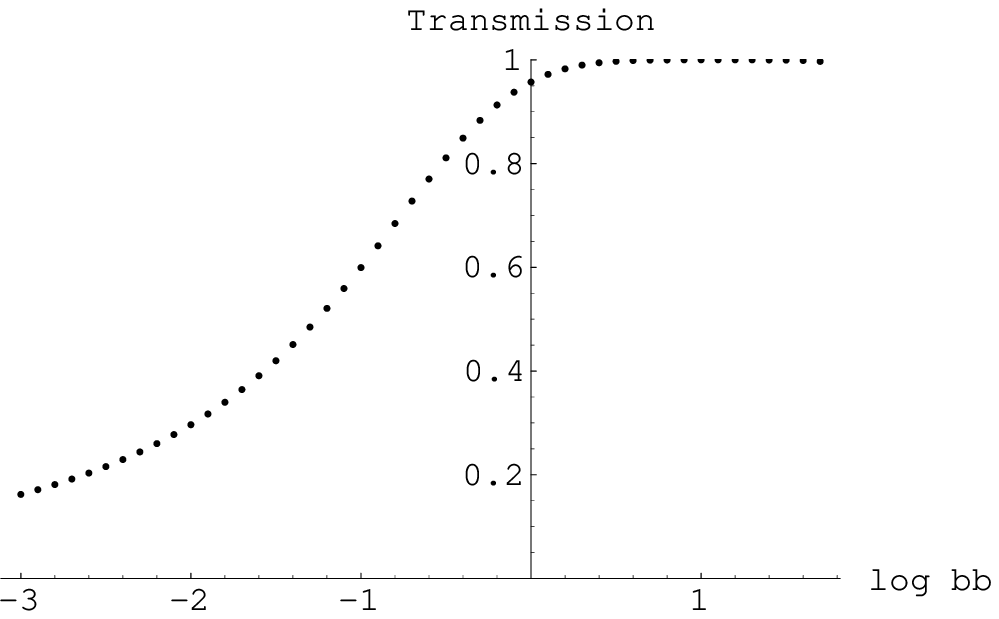,width=0.61\hsize}{Transmission for different
values of the
  deformation $\log c$}
\label{transmission}

For small values of $c$ the transmission vanishes. That is, in the
limit of two branes intersecting on a line the degrees of freedom
on the two branes decouple.  On the other hand, for large $c$,
when the intersection is deformed away, the transmission
approaches $1$ and there is really only one brane left and the
degrees of freedom in the horizontal and vertical asymptotic
regions are connected as one would expect it for a single brane.

\section{Conclusions}

In this note we have given a few examples in which calibration
equations arise as solutions of F and D-flatness conditions when
the action for a brane is written in terms of a lower dimensional
superspace.  In fact this result is quite general and should
include various calibration conditions which we have not mentioned
here.

The lower dimensional superspace formalism is particularly useful
to write the action for intersecting branes, as it facilitates
writing the couplings between ambient fields and the twist fields
localized at the intersections. We have explicitly shown how
intersecting branes recombine into smooth calibrated manifolds
when the twist fields condense on the Higgs branch.

In the case of special Lagrangian manifolds,  the Yang Mills
description suffices to give the Lagrangian condition, arising
from the F-term in the appropriate lower dimensional superspace. A
linearized special-Lagrangian condition arises when D-flatness is
also considered. The non-linear terms vanish for the special case
of a holomorphic curve times a line,  which we have shown to arise
on the Higgs branch of the gauge theory describing a triple
intersection of D-branes.   It would be interesting to obtain the
full non-linear special Lagrangian condition from a D-flatness
condition in supersymmetric version of the Dirac-Born-Infeld
theory.


\acknowledgments
We would like to thank Dieter L\"ust for discussions which stimulated
this investigation. Moreover
we benefited a lot from discussions with Vijay Balasubramanian, David
Berman, Gary Gibbons, Michael Green, Stephan Stieberger and
David Tong. J.E., Z.G. and I.K. are supported by the
DFG (Deutsche Forschungsgemeinschaft) within the `Emmy Noether'
Programme, grant ER301/1--2. R.C.H.~is supported by the
TMR Network ``Supertringtheory'', EU contract HPRN-CT-2000-00122.

\bibliographystyle{JHEP}
\bibliography{triple}

\providecommand{\href}[2]{#2}\begingroup\raggedright\begin{thebibliography}{10}

\bibitem{Smith}
D.~J. Smith, {\it Intersecting brane solutions in string and m-theory},  {\em
  Class. Quant. Grav.} {\bf 20} (2003) R233,
  [\href{http://xxx.lanl.gov/abs/hep-th/0210157}{{\tt hep-th/0210157}}].

\bibitem{MSS}
N.~Marcus, A.~Sagnotti, and W.~Siegel, {\it Ten-dimensional supersymmetric
  yang-mills theory in terms of four-dimensional superfields},  {\em Nucl.
  Phys.} {\bf B224} (1983) 159.

\bibitem{AHGW}
N.~Arkani-Hamed, T.~Gregoire, and J.~Wacker, {\it Higher dimensional
  supersymmetry in 4d superspace},  {\em JHEP} {\bf 03} (2002) 055,
  [\href{http://xxx.lanl.gov/abs/hep-th/0101233}{{\tt hep-th/0101233}}].

\bibitem{EGK}
J.~Erdmenger, Z.~Guralnik, and I.~Kirsch, {\it Four-dimensional superconformal
  theories with interacting boundaries or defects},  {\em Phys. Rev.} {\bf D66}
  (2002) 025020, [\href{http://xxx.lanl.gov/abs/hep-th/0203020}{{\tt
  hep-th/0203020}}].

\bibitem{CEGK1}
N.~R. Constable, J.~Erdmenger, Z.~Guralnik, and I.~Kirsch, {\it Intersecting
  d3-branes and holography},
  \href{http://xxx.lanl.gov/abs/hep-th/0211222}{{\tt hep-th/0211222}}.

\bibitem{CEGK2}
N.~R. Constable, J.~Erdmenger, Z.~Guralnik, and I.~Kirsch, {\it
  (de)constructing intersecting m5-branes},  {\em Phys. Rev.} {\bf D67} (2003)
  106005, [\href{http://xxx.lanl.gov/abs/hep-th/0212136}{{\tt
  hep-th/0212136}}].

\bibitem{CEGK3}
N.~R. Constable, J.~Erdmenger, Z.~Guralnik, and I.~Kirsch, {\it Intersecting
  branes, defect conformal field theories and tensionless strings},  {\em
  Fortsch. Phys.} {\bf 51} (2003) 732--737,
  [\href{http://xxx.lanl.gov/abs/hep-th/0212265}{{\tt hep-th/0212265}}].

\bibitem{AW}
M.~Atiyah and E.~Witten, {\it M-theory dynamics on a manifold of g(2)
  holonomy},  {\em Adv. Theor. Math. Phys.} {\bf 6} (2003) 1--106,
  [\href{http://xxx.lanl.gov/abs/hep-th/0107177}{{\tt hep-th/0107177}}].

\bibitem{GT}
S.~Gukov and D.~Tong, {\it D-brane probes of special holonomy manifolds, and
  dynamics of n = 1 three-dimensional gauge theories},  {\em JHEP} {\bf 04}
  (2002) 050, [\href{http://xxx.lanl.gov/abs/hep-th/0202126}{{\tt
  hep-th/0202126}}].

\bibitem{BGKL1}
R.~Blumenhagen, L.~Goerlich, B.~K{\"o}rs, and D.~L{\"u}st, {\it Noncommutative
  compactifications of type i strings on tori with magnetic background flux},
  {\em JHEP} {\bf 10} (2000) 006,
  [\href{http://xxx.lanl.gov/abs/hep-th/0007024}{{\tt hep-th/0007024}}].

\bibitem{BGKL2}
R.~Blumenhagen, L.~Goerlich, B.~K{\"o}rs, and D.~L{\"u}st, {\it Magnetic flux
  in toroidal type i compactification},  {\em Fortsch. Phys.} {\bf 49} (2001)
  591--598, [\href{http://xxx.lanl.gov/abs/hep-th/0010198}{{\tt
  hep-th/0010198}}].

\bibitem{AFIRU}
G.~Aldazabal, S.~Franco, L.~E. Ibanez, R.~Rabadan, and A.~M. Uranga, {\it
  Intersecting brane worlds},  {\em JHEP} {\bf 02} (2001) 047,
  [\href{http://xxx.lanl.gov/abs/hep-ph/0011132}{{\tt hep-ph/0011132}}].

\bibitem{IMR}
L.~E. Ibanez, F.~Marchesano, and R.~Rabadan, {\it Getting just the standard
  model at intersecting branes},  {\em JHEP} {\bf 11} (2001) 002,
  [\href{http://xxx.lanl.gov/abs/hep-th/0105155}{{\tt hep-th/0105155}}].

\bibitem{K1}
C.~Kokorelis, {\it Exact standard model compactifications from intersecting
  branes},  {\em JHEP} {\bf 08} (2002) 036,
  [\href{http://xxx.lanl.gov/abs/hep-th/0206108}{{\tt hep-th/0206108}}].

\bibitem{K2}
C.~Kokorelis, {\it New standard model vacua from intersecting branes},  {\em
  JHEP} {\bf 09} (2002) 029,
  [\href{http://xxx.lanl.gov/abs/hep-th/0205147}{{\tt hep-th/0205147}}].

\bibitem{K3}
C.~Kokorelis, {\it Exact standard model structures from intersecting d5-
  branes},  \href{http://xxx.lanl.gov/abs/hep-th/0207234}{{\tt
  hep-th/0207234}}.

\bibitem{Ibanez}
D.~Cremades, L.~E. Ibanez, and F.~Marchesano, {\it Intersecting brane models of
  particle physics and the higgs mechanism},  {\em JHEP} {\bf 07} (2002) 022,
  [\href{http://xxx.lanl.gov/abs/hep-th/0203160}{{\tt hep-th/0203160}}].

\bibitem{Hashimoto}
K.~Hashimoto and S.~Nagaoka, {\it Recombination of intersecting d-branes by
  local tachyon condensation},  {\em JHEP} {\bf 06} (2003) 034,
  [\href{http://xxx.lanl.gov/abs/hep-th/0303204}{{\tt hep-th/0303204}}].

\bibitem{Huang}
W.-H. Huang, {\it Recombination of intersecting d-branes in tachyon field
  theory},  {\em Phys. Lett.} {\bf B564} (2003) 155--162,
  [\href{http://xxx.lanl.gov/abs/hep-th/0304171}{{\tt hep-th/0304171}}].

\bibitem{HT}
K.~Hashimoto and W.~Taylor, {\it Strings between branes},
  \href{http://xxx.lanl.gov/abs/hep-th/0307297}{{\tt hep-th/0307297}}.

\bibitem{JT}
N.~T. Jones and S.~H.~H. Tye, {\it Spectral flow and boundary string field
  theory for angled d-branes},  {\em JHEP} {\bf 08} (2003) 037,
  [\href{http://xxx.lanl.gov/abs/hep-th/0307092}{{\tt hep-th/0307092}}].

\bibitem{Khoury:2001wf}
J.~Khoury, B.~A. Ovrut, P.~J. Steinhardt, and N.~Turok, {\it The ekpyrotic
  universe: Colliding branes and the origin of the hot big bang},  {\em Phys.
  Rev.} {\bf D64} (2001) 123522,
  [\href{http://xxx.lanl.gov/abs/hep-th/0103239}{{\tt hep-th/0103239}}].

\bibitem{ST}
P.~J. Steinhardt and N.~Turok, {\it A cyclic model of the universe},  {\em
  Science} {\bf 296} (2002) 1436--1439.

\bibitem{S}
S.~Sethi, {\it The matrix formulation of type iib five-branes},  {\em Nucl.
  Phys.} {\bf B523} (1998) 158--170,
  [\href{http://xxx.lanl.gov/abs/hep-th/9710005}{{\tt hep-th/9710005}}].

\bibitem{GS}
O.~J. Ganor and S.~Sethi, {\it New perspectives on yang-mills theories with
  sixteen supersymmetries},  {\em JHEP} {\bf 01} (1998) 007,
  [\href{http://xxx.lanl.gov/abs/hep-th/9712071}{{\tt hep-th/9712071}}].

\bibitem{KS}
A.~Kapustin and S.~Sethi, {\it The higgs branch of impurity theories},  {\em
  Adv. Theor. Math. Phys.} {\bf 2} (1998) 571--591,
  [\href{http://xxx.lanl.gov/abs/hep-th/9804027}{{\tt hep-th/9804027}}].

\bibitem{KR}
A.~Karch and L.~Randall, {\it Open and closed string interpretation of susy
  cft's on branes with boundaries},  {\em JHEP} {\bf 06} (2001) 063,
  [\href{http://xxx.lanl.gov/abs/hep-th/0105132}{{\tt hep-th/0105132}}].

\bibitem{DFO}
O.~DeWolfe, D.~Z. Freedman, and H.~Ooguri, {\it Holography and defect conformal
  field theories},  {\em Phys. Rev.} {\bf D66} (2002) 025009,
  [\href{http://xxx.lanl.gov/abs/hep-th/0111135}{{\tt hep-th/0111135}}].

\bibitem{AdWFK}
O.~Aharony, O.~DeWolfe, D.~Z. Freedman, and A.~Karch, {\it Defect conformal
  field theory and locally localized gravity},  {\em JHEP} {\bf 07} (2003) 030,
  [\href{http://xxx.lanl.gov/abs/hep-th/0303249}{{\tt hep-th/0303249}}].

\bibitem{BGK}
A.~Bergman, O.~J. Ganor, and J.~L. Karczmarek, {\it A note on intersecting and
  fluctuating solitons in 4d noncommutative field theory},  {\em Phys. Rev.}
  {\bf D64} (2001) 065001, [\href{http://xxx.lanl.gov/abs/hep-th/0101095}{{\tt
  hep-th/0101095}}].

\bibitem{MvR}
M.~Van~Raamsdonk, {\it Blending local symmetries with matrix nonlocality in d-
  brane effective actions},  \href{http://xxx.lanl.gov/abs/hep-th/0305145}{{\tt
  hep-th/0305145}}.

\bibitem{KMG}
S.~Kachru and J.~McGreevy, {\it Supersymmetric three-cycles and (super)symmetry
  breaking},  {\em Phys. Rev.} {\bf D61} (2000) 026001,
  [\href{http://xxx.lanl.gov/abs/hep-th/9908135}{{\tt hep-th/9908135}}].

\bibitem{BBH}
R.~Blumenhagen, V.~Braun, and R.~Helling, {\it Bound states of d(2p)-d0 systems
  and supersymmetric p- cycles},  {\em Phys. Lett.} {\bf B510} (2001) 311--319,
  [\href{http://xxx.lanl.gov/abs/hep-th/0012157}{{\tt hep-th/0012157}}].

\bibitem{CK}
J.~L. Karczmarek and J.~Callan, Curtis~G., {\it Tilting the noncommutative
  bion},  {\em JHEP} {\bf 05} (2002) 038,
  [\href{http://xxx.lanl.gov/abs/hep-th/0111133}{{\tt hep-th/0111133}}].

\bibitem{Joyce}
D.~Joyce, {\it Lectures on special lagrangian geometry},
  \href{http://xxx.lanl.gov/abs/math.dg/0111111}{{\tt math.dg/0111111}}.

\bibitem{BBS}
K.~Becker, M.~Becker, and A.~Strominger, {\it Five-branes, membranes and
  nonperturbative string theory},  {\em Nucl. Phys.} {\bf B456} (1995)
  130--152, [\href{http://xxx.lanl.gov/abs/hep-th/9507158}{{\tt
  hep-th/9507158}}].

\bibitem{FF}
J.~M. Figueroa-O'Farrill, {\it Intersecting brane geometries},  {\em J. Geom.
  Phys.} {\bf 35} (2000) 99--125,
  [\href{http://xxx.lanl.gov/abs/hep-th/9806040}{{\tt hep-th/9806040}}].

\bibitem{KLM}
A.~Karch, D.~L{\"u}st, and A.~Miemiec, {\it N = 1 supersymmetric gauge theories
  and supersymmetric 3- cycles},  {\em Nucl. Phys.} {\bf B553} (1999) 483--510,
  [\href{http://xxx.lanl.gov/abs/hep-th/9810254}{{\tt hep-th/9810254}}].

\bibitem{Ketov}
S.~V. Ketov, {\it Many faces of born-infeld theory},
  \href{http://xxx.lanl.gov/abs/hep-th/0108189}{{\tt hep-th/0108189}}.

\bibitem{Gibbons}
G.~W. Gibbons, {\it Branes as bions},  {\em Class. Quant. Grav.} {\bf 16}
  (1999) 1471--1477, [\href{http://xxx.lanl.gov/abs/hep-th/9803203}{{\tt
  hep-th/9803203}}].

\bibitem{Townsend}
P.~K. Townsend, {\it Brane theory solitons},
  \href{http://xxx.lanl.gov/abs/hep-th/0004039}{{\tt hep-th/0004039}}.

\bibitem{OV}
H.~Ooguri and C.~Vafa, {\it Knot invariants and topological strings},  {\em
  Nucl. Phys.} {\bf B577} (2000) 419--438,
  [\href{http://xxx.lanl.gov/abs/hep-th/9912123}{{\tt hep-th/9912123}}].

\bibitem{Abel}
S.~A. Abel and A.~W. Owen, {\it Interactions in intersecting brane models},
  {\em Nucl. Phys.} {\bf B663} (2003) 197--214,
  [\href{http://xxx.lanl.gov/abs/hep-th/0303124}{{\tt hep-th/0303124}}].

\bibitem{CP}
M.~Cvetic and I.~Papadimitriou, {\it Conformal field theory couplings for
  intersecting d-branes on orientifolds},  {\em Phys. Rev.} {\bf D68} (2003)
  046001, [\href{http://xxx.lanl.gov/abs/hep-th/0303083}{{\tt
  hep-th/0303083}}].

\bibitem{CIM}
D.~Cremades, L.~E. Ibanez, and F.~Marchesano, {\it Yukawa couplings in
  intersecting d-brane models},  {\em JHEP} {\bf 07} (2003) 038,
  [\href{http://xxx.lanl.gov/abs/hep-th/0302105}{{\tt hep-th/0302105}}].

\bibitem{Lambert}
N.~D. Lambert, {\it Moduli and brane intersections},  {\em Phys. Rev.} {\bf
  D67} (2003) 026006, [\href{http://xxx.lanl.gov/abs/hep-th/0209141}{{\tt
  hep-th/0209141}}].

\bibitem{Sato}
T.~Sato, {\it A quantum analysis on recombination process and dynamics of
  d-p-branes at one angle},  \href{http://xxx.lanl.gov/abs/hep-th/0304237}{{\tt
  hep-th/0304237}}.

\bibitem{CM}
J.~Callan, Curtis~G. and J.~M. Maldacena, {\it Brane dynamics from the
  born-infeld action},  {\em Nucl. Phys.} {\bf B513} (1998) 198--212,
  [\href{http://xxx.lanl.gov/abs/hep-th/9708147}{{\tt hep-th/9708147}}].

\end{thebibliography}\endgroup

\end{document}